\documentclass[%
twocolumn, 
superscriptaddress,
showpacs,
 amsmath,amssymb,
  prl,
nofootinbib,
]{revtex4-1}
\usepackage{graphicx}
\usepackage{dcolumn}
\usepackage{bm}

\usepackage{dcolumn}

\newlength{\latticesep}
\def \beq {\begin{eqnarray}}
\def \eeq {\end{eqnarray}}
\def \bfr {{\bf r}}
\def \Schrodinger {{Schr\"{o}dinger }}
\def \rbar  {{\bar{r}}}

\begin{document}


\title{Similarity transformation of the electronic \Schrodinger equation via Jastrow factorisation}                

\author{Aron J. Cohen}
\email{a.cohen@fkf.mpg.de}
\affiliation{%
Max Planck Institute for Solid State Research, Heisenbergstr. 1, 70569 Stuttgart, Germany
}
\author{Hongjun Luo}%
\affiliation{%
Max Planck Institute for Solid State Research, Heisenbergstr. 1, 70569 Stuttgart, Germany
}
\author{Kai Guther}
\affiliation{%
Max Planck Institute for Solid State Research, Heisenbergstr. 1, 70569 Stuttgart, Germany
}
\author{Werner Dobrautz}
\affiliation{%
Max Planck Institute for Solid State Research, Heisenbergstr. 1, 70569 Stuttgart, Germany
}
\author{David P. Tew}
\affiliation{%
Max Planck Institute for Solid State Research, Heisenbergstr. 1, 70569 Stuttgart, Germany
}
\author{Ali Alavi}

\email{a.alavi@fkf.mpg.de}

\affiliation{%
Max Planck Institute for Solid State Research, Heisenbergstr. 1, 70569 Stuttgart, Germany
}%
\affiliation{
 Dept of Chemistry, University of Cambridge, Lensfield Road, Cambridge CB2 1EW, United Kingdom
}%

\date{\today}

\begin{abstract}
By expressing the electronic wavefunction in an explicitly-correlated (Jastrow-factorised) form, a similarity-transformed effective 
Hamiltonian can be derived. The effective Hamiltonian is non-Hermitian and contains three-body interactions. The resulting ground-state eigenvalue problem can be solved projectively using a stochastic configuration-interaction formalism. Our approach permits use of highly flexible Jastrow functions, which we show to be effective in achieving  
extremely high accuracy, even with small basis sets.   
Results are presented for the total energies and ionisation potentials of the first-row atoms, achieving accuracy within a mH 
of the basis-set limit, using modest basis sets and computational effort.    
\end{abstract}


\pacs{Valid PACS appear here}
\maketitle


Methods aiming to obtain high-accuracy solutions to the electronic \Schrodinger equation must tackle two essential components of the problem, 
namely providing highly flexible 
expansions capable of resolving non-analytic features of the wavefunction, including the Kato cusps\cite{Kato:CPAM10-151} at electron coalesence points, 
as well as treatment of many-electron correlation at medium and long range.  The combination of these
two facets of the problem leads to overwhelming computational complexity, requiring large basis sets and high-order correlation methods, 
approximations to which can result in a significant loss of accuracy. The goal of achieving 
``chemical" accuracy remains extremely challenging for all but the simplest systems.   

In Fock space approaches, including the majority of quantum chemical methodologies based on configurational expansions, the first-quantised \Schrodinger Hamiltonian is replaced by a second-quantised 
form, expressed in a one-electron basis.  The passage from first quantisation to second is invoked primarily to impose anti-symmetry 
on the solutions, via fermionic creation and annihilation operators of the orbital basis. However, this formulation loses the 
ability to explicitly include {\em electron pair} variables (such as electron-electron distances) into the wavefunction, which has long been known \cite{Hylleraas1929} 
to be crucial in obtaining 
an efficient description of electron correlation. Correlation effects are then indirectly obtained via superpositions of 
Slater determinants over the Fock space, as in configuration interaction, coupled-cluster and tensor-decomposition methods. These are computationally costly methods, especially with large basis sets. In quantum chemistry, explicitly-correlated methods usually proceed via the R12 formalism of Kutzelnigg \cite{Kutzelnigg1985}, and its more modern F12 variants 
\cite{Ten-noS_CPL_2004}, in combination with perturbation theory \cite{KutzelniggKlopper1991} or coupled-cluster theory \cite{NogaJ_JCP_1994}. These methods augment the Fock-space (configurational) wavefunctions with strongly-orthogonal geminal terms with fixed amplitudes, imposing a first-order cusp condition. This approximation is suitable for systems whose ground state wavefunction is dominated by a single determinant. The inclusion of explicit correlation in strongly correlated, multi-determinantal, wavefunctions remains an open challenge. 

In this paper, we postpone the passage to second quantisation until after electron-pair information 
has been incorporated into the wavefunction. This is achieved by factorising the electronic wavefunction $\Psi$ in Jastrow\cite{Jastrow} form:
\beq
  \Psi=e^\tau \Phi   \label{PsiJas}
\eeq 
where $\tau=\sum_{i<j} u(\bfr_i,\bfr_j)$ with $u(\bfr_i,\bfr_j)=u(\bfr_j,\bfr_i)$ is a symmetric correlation function over electron pairs, and $\Phi$ is the associated  
many-body function we will aim to compute. The precise form of $u$ has a significant bearing 
on the efficacy of the method \cite{PhysRev.92.609,PhysRevB.18.3126,RevModPhys.73.33,DrummondND_PRB_2004,PhysRevE.86.036703}, and will be later discussed. Substituting \ref{PsiJas} into the \Schrodinger equation $\hat{H}\Psi=E\Psi$, 
and rearranging, we obtain $\Phi$ as an eigenfunction of the similarity transformed (ST) Hamiltonian $\tilde{H}$, i.e.
$ \tilde{H}\Phi  =  E \Phi$, with 
\beq
  \tilde{H} = e^{-\tau} \hat{H} e^\tau = \hat{H} + [\hat{H},\tau] + \frac{1}{2}[[\hat{H},\tau],\tau] 
\eeq
The commutator expansion truncates at second order because the only terms in $\hat{H}$ which do not commute with $\tau$ are 
the (second-derivative) kinetic energy operators. The explicit form of $\tilde{H}$ contains additional two- and three-body terms:
\begin{eqnarray*}
\tilde{H}  & = &  \hat{H}-\sum_{i}\left(\frac{1}{2}\triangledown_{i}^{2}\tau+(\triangledown_{i}\tau)\triangledown_{i}+\frac{1}{2}(\triangledown_{i}\tau)^{2}\right)\\
 &     = &     \hat{H}-\sum_{i< j}\hat{K}({\bf r}_{i},{\bf r}_{j})-\sum_{i<j<k}\hat{L}({\bf r}_{i},{\bf r}_{j},{\bf r}_{k}).
\end{eqnarray*}
where 
\begin{widetext}
\begin{eqnarray*}
       \hat{K} ({\bf r}_{i},{\bf r}_{j})&=&\frac{1}{2}\left(\triangledown_{i}^{2}u({\bf r}_{i},{\bf r}_{j})+\triangledown_{j}^{2}u({\bf r}_{i},{\bf r}_{j})+(\triangledown_{i}u({\bf r}_{i},{\bf r}_{j}))^{2}+(\triangledown_{j}u({\bf r}_{j},{\bf r}_{i}))^{2}\right)\\
       & & +\left(\triangledown_{i}u({\bf r}_{i},{\bf r}_{j}))\cdot\triangledown_{i}+(\triangledown_{j}u({\bf r}_{i},{\bf r}_{j}))\cdot\triangledown_{j}\right)
     \\
 \hat{L}({\bf r}_{i},{\bf r}_{j},{\bf r}_{k})&=&\triangledown_{i}u({\bf r}_{i},{\bf r}_{j})\cdot \triangledown_{i}u({\bf r}_{i},{\bf r}_{k})+\triangledown_{j}u({\bf r}_{j},{\bf r}_{i})\cdot \triangledown_{j}u({\bf r}_{j},{\bf r}_{k})+
 \triangledown_{k}u({\bf r}_{k},{\bf r}_{i})\cdot \triangledown_{k}u({\bf r}_{k},{\bf r}_{j})
\end{eqnarray*}
\end{widetext}
The similarity transformed Hamiltonian is non-Hermitian, owing to the gradient terms in $\hat{K}$. Projective techniques can be used to obtain the distinct right or left eigenvectors for a given eigenvalue $E$.
The FCIQMC method (full configuration interaction quantum Monte Carlo) and its initiator approximation \cite{original-fciqmc,initiator-fciqmc} has been previously adapted for this purpose \cite{LuoAlavi2018, DobrautzLuoAlavi2019} and we use it in this study. Note that our method differs from the Transcorrelated (TC) Method of Boys and Handy \cite{BoysHandy1969} and modern extensions \cite{Tsuneyuki2008} in three crucial respects: we solve for $\Phi$ as a full multi-determinant expansion (to be obtained via an FCIQMC procedure) whilst $\Phi$ is a single Slater determinant in their work. Second, the formal unitary invariance of our $\Phi$ negates the need for orbital optimisation and we simply use Hartree-Fock orbitals as the basis of our Fock space. Third, we do not attempt simultaneous optimisation of the Jastrow function and $\Phi$. Many of the difficulties associated with the non-Hermitian nature of $\tilde{H}$, which have plagued  many previous attempts at the TC method, are thus avoided. The multi-determinant nature of $\Phi$ also gives much greater flexibility to this function than a single Slater determinant, which we believe to be crucial in obtaining high accuracy. Indeed $\Phi$ must share the same nodal surface as $\Psi$ for an exact factorisation, and a full CI form for $\Phi$ gives it much more flexibility in this regard than orbital optimisation within a Slater-Jastrow form. This is a fundamental advantage of the present method, in addition to the avoidance of the often troublesome redundancy in the orbital optimisation and Jastrow optimisation procedure. Of course the price to be paid is a formally exponential scaling method. However the cost of this can be ameliorated via the stochastic FCIQMC procedure. 
Our approach differs from that of the related work of Ten-no \cite{Ten-noS_CPL_2000a,hino_application_2002} in that we use a highly flexible form for both the Jastrow and multi-determinental  expansion, rather than using a fixed short-ranged Jastrow function and a low order perturbative or coupled-cluster expansion.
\begin{table*}[!t]
\caption{Total atomic energies (Hartrees), for CCSD(T), CCSD(T)-F12, and the ST Hamiltonian, using the SM7, and SM17 correlation factors. MAE for each method across the series is also shown.}
\begin{tabular}{r  r | r   r  r  r  r  r  r   r|  r}    
 method &   basis &        Li &        Be  &         B   &          C &         N &         O  &         F   &         Ne  &   MAE   \\     \hline\hline
CCSD(T) 
&  cc-pVDZ      & -7.43264  &  -14.61741 &  -24.59026  & -37.76156  & -54.47994 & -74.911155 & -99.52932   & -128.68069  &   0.121   \\
&  cc-pVTZ      & -7.44606  &  -14.62379 &  -24.60538  & -37.78953  & -54.52487 & -74.98494  & -99.63219   & -128.81513  &   0.069    \\
&  cc-pVQZ      & -7.44983  &  -14.64008 &  -24.62350  & -37.81209  & -54.55309 & -75.02319  & -99.68158   & -128.87676  &   0.039   \\[1ex]
F12
&  cc-pVDZ      & -7.47458  &  -14.65400 &  -24.63121  & -37.80901  & -54.53707 & -74.99208  & -99.63623   & -128.81125  &   0.053   \\
&  cc-pVTZ      & -7.47267  &  -14.65653 &  -24.63626  & -37.81883  & -54.55293 & -75.01752  & -99.66994   & -128.85890  &   0.036   \\
&  cc-pVQZ      & -7.47370  &  -14.65933 &  -24.64187  & -37.82884  & -54.56916 & -75.04056  & -99.70070   & -128.89816  &   0.020   \\[1ex]
SM7  & cc-pVDZ &  -7.46726 &  -14.65517 &  -24.63279  &  -37.81469 & -54.53448 &  -74.97785 &  -99.60602  & -128.78385  &   0.063               \\       
     & cc-pVTZ &  -7.47627 &  -14.65943 &  -24.64458  &  -37.83703 & -54.57236 &  -75.04055 &  -99.69421  & -128.89389  &   0.019                \\      
     & cc-pVQZ &  -7.47785 &  -14.66791 &  -24.65417  &  -37.84791 & -54.58778 &  -75.06296 &  -99.72507  & -128.92967  &   0.003      \\[1ex]
SM17  &cc-pVDZ &  -7.47707 &  -14.66793 &  -24.64521  & -37.82772  & -54.55719 &  -75.01639 &  -99.65834  & -128.83682  &   0.036      \\
      &cc-pVTZ &  -7.47804 &  -14.66789 &  -24.65003  & -37.83928  & -54.57989 &  -75.05303 &  -99.71377  & -128.90944  &   0.010      \\
      &cc-pVQZ &  -7.47845 &  -14.66749 &  -24.65287  & -37.84461  & -54.58844 &  -75.06609 &  -99.73283  & -128.93542  &   0.001  \\[1ex] 
          Expt & & -7.47806 &  -14.66736 &  -24.65391  & -37.84500  & -54.58920 & -75.06730  & -99.73390   & -128.93760  &             \\   
\end{tabular}
\end{table*}

Using this first-quantised Hamiltonian, we can construct a second-quantised Hamiltonian for a given set of orbitals $\{\phi_1,....,\phi_M\}$, 
with corresponding spin-$\frac{1}{2}$ creation (annihilation) operators  $a^\dagger_{p\sigma}  (a_{p\sigma})$: 
\begin{widetext}
\begin{align}
\tilde{H}=   \sum_{pq\sigma} h^p_q a^\dagger_{p\sigma} a_{q\sigma}+ \frac{1}{2} \sum_{pqrs } (V^{pq}_{rs}-K^{pq}_{rs}) \sum_{\sigma \tau} a^\dagger_{p\sigma} a^\dagger_{q\tau} a_{s\tau} a_{r\sigma} 
  - \frac{1}{6}\sum_{pqrstu} L^{pqr}_{stu}  \sum_{\sigma \tau \lambda} a^\dagger_{p\sigma} a^\dagger_{q\tau} a^\dagger_{r\lambda} a_{u\lambda} a_{t\tau} a_{s\sigma} 
\end{align} 
\end{widetext} 
where $h^p_q=\langle\phi_p|h|\phi_q\rangle$ and $V^{pq}_{rs}=\langle\phi_p\phi_q|r_{12}^{-1}|\phi_r\phi_s\rangle$ are the one- and two-body 
terms of the \Schrodinger Hamiltonian, and $K^{pq}_{rs}=\langle\phi_p\phi_q|\hat{K}|\phi_r\phi_s\rangle$ and $L^{pqr}_{stu}=\langle\phi_p\phi_q\phi_r|\hat{L} |\phi_s\phi_t\phi_u\rangle$  are the corresponding terms arising from the similarity transformation.  Note that the 3-body operator $\hat{L}$ is Hermitian, and for real orbitals has 48-fold symmetry, a useful feature in reducing the memory requirement to store these integrals. 
Nevertheless storage of the 6-index integrals represents the major bottleneck of this methodology, limiting us at present to about 100 orbitals. This bottleneck can be alleviated using tensor-decomposition and fast on-the-fly evaluation of the integrals, and will be the subject of future work.

Although the form of this similarity transformed Hamiltonian has been known for a long time\cite{HirschfelderJO_JCP_1963}, to our knowledge it has never been treated in its full form until now. We retain all three-body terms, motivated in part by our recent study of the two-dimensional Hubbard model using Gutzwiller similatrity transformations \cite{DobrautzLuoAlavi2019}, in which we show that the 3-body terms do not incur a huge cost in the FCIQMC 
formalism, and their full treatment enables essentially exact results to be obtained. Furthermore, this study shows that the similarity transformations can help enormously in the study of strongly correlated systems, by significantly compactifying the right eigenvector of the ground state  (which are generally highly multiconfigurational otherwise). This suggests that the present formalism may also help in treating strongly correlated \textit{ab initio} Hamiltonians in a manner beyond that of post-hoc explicitly correlated methodologies \cite{Torheyden:JCP131-171103}. 
 

The formulation requires the calculation of additional non-Hermitian two-electron matrix elements and Hermitian three-electron matrix elements. These are computed using numerical quadrature over the direct product of
atom centred grids built from Treutler-Ahlrichs radial grids and Lebedev angular grids, obtained from the PySCF program \cite{PYSCF}. The numerical summations factorise into a series of steps with $N_{grid}^{2}N_{bas}^{2}$, $N_{grid}N_{bas}^{4}$ or $N_{grid}N_{bas}^{6}$ cost, each of which are highly parallelisable and convergence of the integrals with grid size is rapid. Further details are provided in the supplementary material. Our numerical approach makes possible use of arbitrary forms of Jastrow function and all integrals necessary to carry out the ST-FCIQMC calculations are readily available.

In the present study of the first-row atoms and cations, we investigated two correlation factors of the form used by Boys and Handy \cite{BoysHandy1969}:
\beq
  u(\bfr_i,\bfr_j) & =&\sum_{mno}^{m+n+o\le 6} c_{mno} (\rbar_{i}^m \rbar_{j}^n+\rbar_{j}^m\rbar_{i}^n)\rbar_{ij}^o,  \quad
   \bar{r}=\frac{r}{1+r} \nonumber
\eeq
This function has electron-electron (e-e), electron-nucleus (e-n), and e-e-n 3-body terms, which arise respectively from terms in which $m=n=0, o>0$, and $m,n>0, o=0$ and $m,n,o>0$ terms. In this form of correlation factor, the $s$-wave Kato cusp condition can be satisfied, but not the $p$-wave or higher\cite{Tew:JCP129-014104}, and this feature ultimately determines the asymptotic rate of convergence to the basis set limit \cite{jcp83rh}. The parameters $c_{mno}$ are taken from the variance-minimisation variational Monte Carlo (VMC) study of the first-row atoms of Schmidt and Moskowitz \cite{SchmidtMoskowitz1990}. 
In one case (termed SM7) we used a form without the e-e-n terms, and in the second (termed SM17) 
we used the full form of the Jastrow factor, with 17 terms, including e-e-n terms. Comparison of the results for the two correlation factors gives insight into the role played by the additional terms in Jastrow factor in the present methodology.

The ST-FCIQMC calculations were performed in standard valence correlation quantum chemical basis sets, cc-pVXZ, X=D,T,Q (with 14, 30 and 55 basis functions respectively). 
The non-Hermitian nature of the Hamiltonian, together with 3-body interactions, have previously been treated in FCIQMC 
\cite{LuoAlavi2018,DobrautzLuoAlavi2019} and implemented in the NECI code \cite{linear-scaling-fciqmc}, and were further adapted for the molecular Hamiltonian presented here.

\begin{figure}[!b]
\includegraphics[scale=0.34,angle=-90]{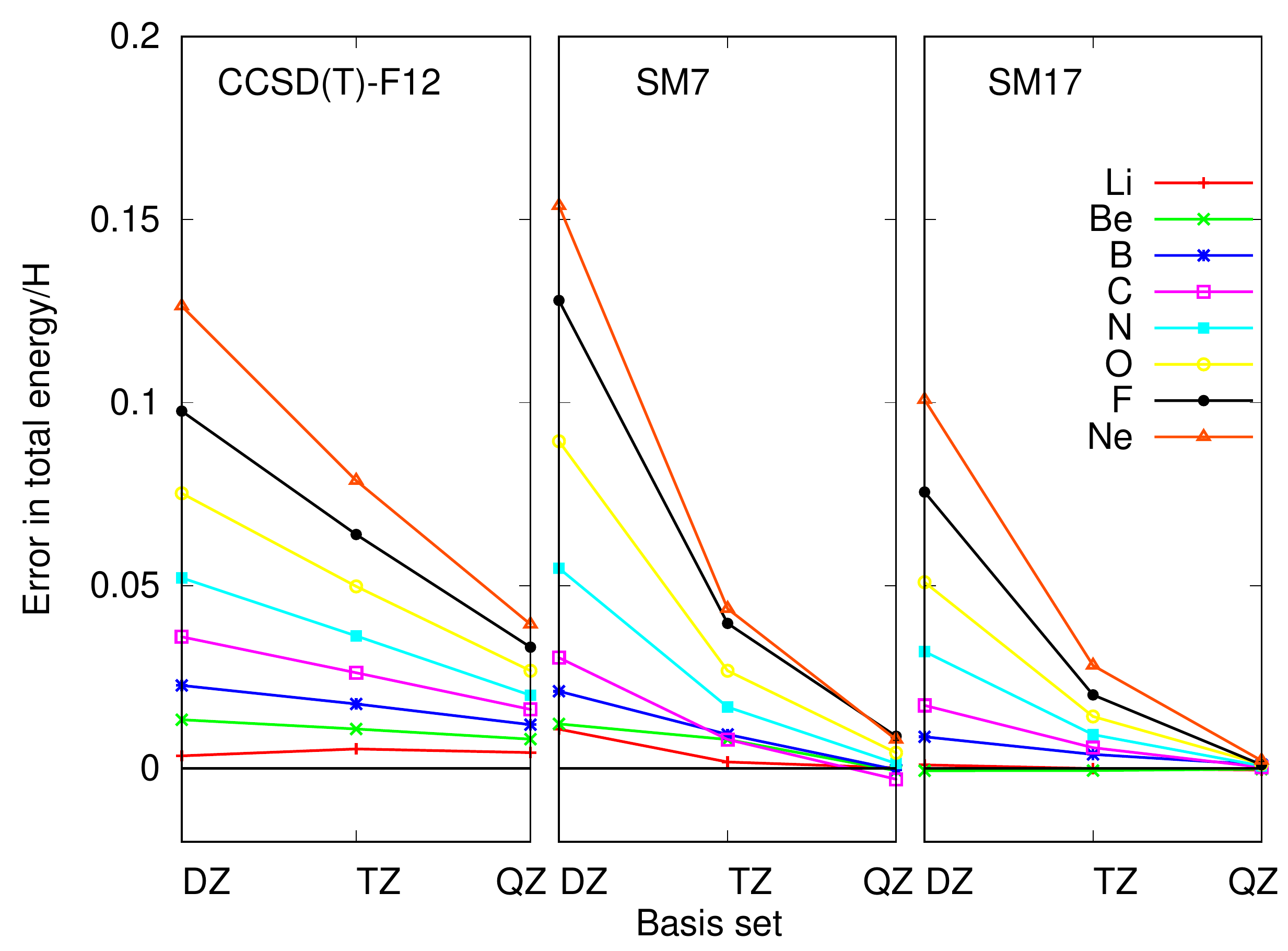}
\caption{
Errors in the total energies of the atoms, in H, for the two correlation functions and the F12 methodology.}
\label{fig:errors_energies}
\end{figure}

\begin{table*}
\caption{IPs in mH, for the CCSD(T)-F12 method, and for the ST Hamiltonian with the SM7 and SM17 correlation factors}
\begin{tabular}{r  r | r   r  r  r  r  r  r   r | r}
method & basis &   Li    &  Be     &  B     &  C      &   N     &  O       &F        &  Ne        &   MAE       \\ \hline\hline 

F12  & cc-pVDZ &  197.70 &  341.36 &  302.47 & 410.24 &  529.60	&  490.44  & 629.71  &  780.26    &   6.07       \\
     & cc-pVTZ &  197.67 &  341.81 &  304.13 & 412.73 &  532.91	&  496.17  & 636.44  &  789.34    &   2.39       \\
     & cc-pVQZ &  197.82 &  341.98 &  304.53 & 413.52 &  534.21	&  498.77  & 639.20  &  792.62    &   0.96       \\[1ex]
SM7  & cc-pVDZ & 195.13  &  341.87 & 297.70 &  404.70 &  522.79 &  474.22  &  617.48 &  768.90    &    13.45     \\                   
     & cc-pVTZ & 198.21  &  342.02 & 303.63 &  412.20 &  531.89 &  491.33  &  631.71 &  785.12    &     4.30     \\                   
     & cc-pVQZ & 198.55  &  342.77 & 304.54 &  413.51 &  533.91 &  497.15  &  637.62 &  790.90    &     1.57     \\[1ex]            
SM17 & cc-pVDZ &  188.50 &  341.40 &  299.54 & 407.46 &  526.84 &  482.62  & 627.60  &  779.22    &     9.65    \\        
     & cc-pVTZ &  198.54 &  342.64 &  305.66 & 414.66 &  535.31 &  499.05  & 640.95  &  793.95    &     0.58      \\       
     & cc-pVQZ &  198.44 &  342.70 &  304.98 & 414.24 &  535.19 &  500.65  & 642.37  &  795.82    &     0.50      \\[1ex]
Expt &       &  198.15 &  342.58 &  304.99 & 413.97 &  534.60 &  500.50  & 641.10  &  794.50    &             \\ 
\end{tabular}
\end{table*}

The results of the ST-FCIQMC calculations for the total atomic energies are shown in Table I. Reference energies are taken from experiment, corrected for relativistic effects as computed by Chakravorty et. al. (1993)\cite{Davidson1993}. The errors in the total energies for the three basis sets are plotted in Fig 1. For comparison with quantum chemistry methods, we also report energies computed using the coupled cluster method CCSD(T)\cite{ccsdt} and its explicitly-correlated variant CCSD(T)-F12\cite{HattigTewKohn2010}. The former gives as indication of the severity of the basis-set problem (for example, a mean absolute error (MAE) of 39mH at cc-pVQZ), whilst the latter show how much this error can be reduced using a state of the art explicit-correlation method when using valence basis sets for all-electron energies (MAE of 20mH at cc-pVQZ).
It is clear that the present methodology gives a marked improvement in the total energies, 
especially using the SM17 correlation factor: an 
MAE of only 1 mH with the cc-pVQZ basis set. A very small degree of non-variationality (less than 1 mH) is observed in a few cases with the lighter elements. Strict variationality is lost in a non-Hermitian formulation, and is often the major concern in Transcorrelated methods, leading to energies far below the exact energy. Here, with the full treatment of ST Hamiltonian, coupled with the projective eigensolver, the results show that this is of no serious concern. The cause of the present non-variationality lies in the fact that the correlation factor does not fulfil the $p$-wave cusp condition, leading to an over-correlation of same-spin pairs of electrons. which prevents the accuracy reaching that obtainable through F12 theory with high-order coupled cluster methods \cite{Klopper2010}. Spin-dependent correlation factors may be a way forward, but incur other complications, and is the subject of current work.

The additional 2 and 3-body terms turn out to make large but generally opposing changes to the total energies. For example, for the Ne atom in the cc-pVQZ basis, with the SM17 correlation factor the expectation value of these terms for Hartree-Fock (HF) determinant are  $\langle D_\text{HF}|\hat{K}|D_\text{HF}\rangle=-382$ mH and $\langle D_\text{HF}|L|D_\text{HF}\rangle=+109$ mH. The effect on the FCI energy is similar: the $\hat{K}$ terms reduce the energy below the exact energy (by 51 mH), whilst the $L$ terms are substantially positive, making the total energy of $\tilde{H}$ exact to within 2 mH. We also see that the SM17 correlation factor is much more effective than SM7. A key property of the SM17 form is that the correlation hole depth can vary depending on the distances of the pair of electrons from the nucleus, getting deeper if the pair are further away from the nucleus. This additional flexibility is very helpful in differentiating between core-electron and valence-electron correlation. 


We obtained the ionisation potentials (IPs) by computing the cation total energies (Table II). Here we used the same Jastrow
parameters as used in the atomic calculations, without further optimisation - this provides a stern test of the  
transferability of methodology. The results show that at both cc-pVTZ and cc-pVQZ basis sets, the MAE for the IPs are only 0.58 and 
0.50mH for the SM17 correlation factor, compared to 2.39 and 0.96 for the CCSD(T)-F12 method. 
The marked improvement of the SM17 results over SM7 highlights the effectiveness of the e-e-n terms in the Jastrow functions to deliver very high accuracy even using the comparatively modest cc-pVTZ basis.

A promising aspect of the present approach is its ability to describe core electron correlation without the need to include tight functions in the basis-set, this already being evident in the excellent total energies of Table I. To investigate this further, in Table III we report the series of the total energy of Neon with differing numbers of electrons, from Ne to Ne$^{7+}$, 
which are increasingly dominated by the core electrons. 
The results from the ST-FCIQMC with the atomic SM17 Jastrow factor are particularly interesting as they give agreement in the total
energy to within a couple of mH for all systems. This is without a core-correlation basis set.
For the other methods such as CCSD(T)-F12, 
a core-valence basis set (e.g. cc-pCVQZ) is essential to describe the total energy. 
This however leads to a very significant increase in the size of the basis set 
(e.g. cc-pVQZ has 55, cc-pCVQZ 84, and cc-pCV5Z 145 basis functions). The ability of the present methodology to capture core correlation via the Jastrow factor, obviating the need to correlate them in the configurational expansion, is a major advantage that will prove even more useful in heavier systems. 

\begin{table*}
\caption{Energies of the cations of Ne using SM17 vs CCSD(T) and CCSD(T)-F12 }
\begin{tabular}{r  r  | r  r  r  r  r  r  r  r }   
 method       &  basis & Ne$^{7+}$ & Ne$^{6+}$ & Ne$^{5+}$ & Ne$^{4+}$ & Ne$^{3+}$ & Ne$^{2+}$ & Ne$^{+}$  & Ne        \\ \hline\hline

 CCSD(T)      & cc-pVQZ   & -102.6530 & -110.2577 & -116.0512 & -120.6884 & -124.2615 & -126.5857 & -128.0871 & -128.8768 \\
 CCSD(T)      & cc-pV5Z   & -102.6585 & -110.2646 & -116.0616 & -120.7011 & -124.2757 & -126.6027 & -128.1067 & -128.8989 \\ 
 CCSD(T)      & cc-pCVQZ  & -102.6788 & -110.2859 & -116.0820 & -120.7209 & -124.2951 & -126.6205 & -128.1224 & -128.9123 \\
 CCSD(T)      & cc-pCV5Z  & -102.6809 & -110.2888 & -116.0871 & -120.7275 & -124.3027 & -126.6303 & -128.1346 & -128.9269 \\
 CCSD(T)-F12  & cc-pCV5Z  & -102.6818 & -110.2900 & -116.0891 & -120.7303 & -124.3062 & -126.6359 & -128.1420 & -128.9360 \\
 ST-FCIQMC    & cc-pVQZ   & -102.6817 & -110.2910 & -116.0886 & -120.7288 & -124.3045 & -126.6334 & -128.1397 & -128.9355 \\ [1ex]

     Expt\cite{Davidson1993}    &        & -102.6822 & -110.2909 & -116.0902 & -120.7312 & -124.3068 & -126.6366 & -128.1431 & -128.9376 \\ 
\end{tabular}
\end{table*}

In closing this section, it is worthwhile re-emphasising a crucial difference between the present method and the standard explicitly correlated methods such as F12, namely how the redundancy between the Jastrow function and the configurational function is dealt with. In F12 methods, strong orthogonality projectors are used to eliminate this redundancy, i.e. correlation that can be described by the Fock space wavefunction (in a given basis set) 
is removed from the correlation factor. As such, only simple correlation functions, typically of the exponential form $u(\bfr_1,\bfr_2)=-\gamma^{-1}e^{-\gamma r_{12}}$ are employed in F12 methods, more complicated forms (such as SM7 and SM17) being less effective because of the projections. In the present method, on the other hand, the configurational function $\Phi$ is explicitly solved {\em in the presence of the potential terms arising from the correlation function}, and can benefit from it: a more realistic correlation function leads to a simpler $\Phi$, and more rapid convergence with respect to the parameters (basis set, CI expansion, etc) that define $\Phi$. Hence the observed significant improvement in performance in going from SM7 to SM17. 

To summarise, we show that eigenfunctions of a Jastrow-factorised similarity-transformed Hamiltonian can be computed using the FCIQMC technique and leads to accurate results for atoms, close to the basis-set limit, even when the configurational wavefunction is expanded in limited basis sets. A major advantage of the present approach, as compared to the F12 methodologies, is that forms of correlation factors beyond pure e-e functions can be used, without the need for projection operators, and deliver excellent energies without the need for augmented basis sets. A further advantage, which comes from the FCI formulation presented, is the ability to tackle strongly correlated systems, such as stretched open-shell molecules. This will be the subject of future work along with the optimisation of Jastrow factors using different trial wavefunctions. The main bottleneck in the current implementation is the need to store 3-body integrals. Work is underway to to alleviate this.


See supplementary material for details of the numerical evaluation of the hamiltonian matrix elements.

The authors gratefully acknowledge funding from the Max Planck Society.



%

\onecolumngrid
\setcounter{equation}{0}
\renewcommand{\theequation}{S\arabic{equation}}

\section{Numerical evaluation of the Hamiltonian Matrix elements}

Our second-quantised similarity-transformed effective Hamiltonian for a given set of orbitals $\{\phi_1,....,\phi_M\}$, 
with corresponding spin-$\frac{1}{2}$ creation (annihilation) operators  $a^\dagger_{p\sigma}  (a_{p\sigma})$ is: 
\begin{align}
\tilde{H}=   \sum_{pq\sigma} h^p_q a^\dagger_{p\sigma} a_{q\sigma}+ \frac{1}{2} \sum_{pqrs } (V^{pq}_{rs}-K^{pq}_{rs}) \sum_{\sigma \tau} a^\dagger_{p\sigma} a^\dagger_{q\tau} a_{s\tau} a_{r\sigma} 
  - \frac{1}{6}\sum_{pqrstu} L^{pqr}_{stu}  \sum_{\sigma \tau \lambda} a^\dagger_{p\sigma} a^\dagger_{q\tau} a^\dagger_{r\lambda} a_{u\lambda} a_{t\tau} a_{s\sigma} 
\end{align} 
where $h^p_q=\langle\phi_p|h|\phi_q\rangle$ and $V^{pq}_{rs}=\langle\phi_p\phi_q|r_{12}^{-1}|\phi_r\phi_s\rangle$ are the one- and two-body 
terms of the \Schrodinger Hamiltonian, and $K^{pq}_{rs}=\langle\phi_p\phi_q|\hat{K}|\phi_r\phi_s\rangle$ and $L^{pqr}_{stu}=\langle\phi_p\phi_q\phi_r|L |\phi_s\phi_t\phi_u\rangle$  are the corresponding terms arising from the similarity transformation.  Note that the 3-body operator $L$ is Hermitian, and for real orbitals has 48-fold symmetry, a useful feature in reducing the memory requirement to store these integrals.

Our formulation requires the calculation of additional non-Hermitian two-electron matrix elements and Hermitian three-electron matrix elements. The integrals are calculated numerically as we outline below.
The additional two-electron matrix elements are,
\begin{eqnarray}
K^{pq(1)}_{rs} &=& \langle pq|\nabla_1u(\bfr_1,\bfr_2).\nabla_1|rs\rangle\\
K^{pq(2)}_{rs} &=& \langle pq|\nabla_1^2u(\bfr_1,\bfr_2)|rs\rangle \\
K^{pq(3)}_{rs} &=& \langle pq|(\nabla_1u(\bfr_1,\bfr_2))^2|rs\rangle
\end{eqnarray}


To calculate these matrix elements we use a simple six-dimensional numerical quadrature by taking a repeated
three-dimensional grid with $N_{grid}$ grid-points and the corresponding quadrature weights $wt$. For example
\begin{equation}
K^{pq(1)}_{rs} = \sum_{m_1}^{N_{grid}}\sum_{m_2}^{N_{grid}}\phi_{p}(\bfr_{m_1})\nabla_{\bfr_{m_1}}\phi_r(\bfr_{m_1}).\nabla_{\bfr_{m_1}}u(\bfr_{m_1},\bfr_{m_2})\phi_{q}({\bf r}_{m_2})\phi_{s}({\bf r}_{m_2})wt(\bfr_{m_1})wt(\bfr_{m_2})
\end{equation}

For the three-dimensional grid we use atom centred grids built from Treutler-Ahlrichs radial grids and Lebedev angular grids from  the DFT quadrature grids from PySCF \cite{PYSCF}.
Naively carried out this would be $N_{grid}^{2}N_{bas}^{4}$ to calculate
the $K$ integrals but this can simply be divided into two steps by
carrying out the sum over ${\bf r}'$ first at each ${\bf r}$, in $N_{grid}^{2}N_{bas}^{2}$ steos, and a final sum costing $N_{grid}N_{bas}^{4}$ steps.
The numerical summation is highly parallelizable, and can be computed
easily to give the integrals to carry out the ST FCIQMC
calculations. All the integrals converge rapidly with respect to grid size
except for $K^{pq(2)}_{rs}$,
which is much more efficiently calculated using integration by parts,
\begin{equation*}
K^{pq(2)}_{rs}=-\int\int\phi_{p}({\bf r}_{2})\phi_{r}({\bf r}_{2})\nabla_{1}u({\bf r}_{1},{\bf r}_{2})\ldotp\left[\nabla\phi_{q}({\bf r}_{1})\phi_{s}({\bf r}_{1})+\phi_{q}({\bf r}_{1})\nabla\phi_{s}({\bf r}_{1})\right]{\rm d}{\bf r}_{1}{\rm d}{\bf r}_{2}
\end{equation*}

Another advantage of a grid based representation is that the integrals can be directly carried out in the MO basis. 
This requires only the MO values and derivatives at each grid point, 
$\left\{ \phi_{p}(\bfr_{m}),\nabla\phi_{p}(\bfr_{m})\right\}$, and negates the need for any costly integral transformations.

The three electron matrix elements,
\begin{equation}
L^{pqr(123)}_{stu} = \langle pqr|\nabla_1u(\bfr_1,\bfr_2).\nabla_1u(\bfr_1,\bfr_3)|stu\rangle
\end{equation}
are similarly calculated using numerical methods.
We first create an intermediate vector of $\bfr_1$ by integrating over the other electron first
\begin{equation}
{\bf V}_{qt}(\bfr_1) = \int\phi_q(\bfr_2) \nabla_1 u(\bfr_1,\bfr_2) \phi_t(\bfr_2){\rm d}\bfr_2
\end{equation}
Thus the full three-electron matrix element can be evaluated by,
\begin{equation}
L^{pqr(123)}_{stu} = \int \phi_p(\bfr_1) {\bf V}_{qt}(\bfr_1).{\bf V}_{ru}(\bfr_1)\phi_s(\bfr_1){\rm d}\bfr_1
\end{equation}
The first step is order $N_{grid}^2N_{bas}{^2}$ and the second step is $N^{grid}N_{bas}^6$ but again is simply parallelizable.
To use the full symmetry of the L matrix, which is hermitian, we calculate $L^{pqr123}_{stu}+L^{pqr312}_{stu}+L^{pqr231}_{stu}$ and only calculate for the unique indices $p\ge s, q\ge t,r \ge u, ps\ge qt\ge ru$ and $pqr\ge stu$.


\end{document}